\documentclass[a4paper]{jpconf}
\usepackage{graphicx}
\usepackage{amssymb}

\def\slashchar#1{\setbox0=\hbox{$#1$}
   \dimen0=\wd0 \setbox1=\hbox{/} \dimen1=\wd1
   \ifdim\dimen0>\dimen1 \rlap{\hbox to \dimen0{\hfil/\hfil}} #1
   \else  \rlap{\hbox to \dimen1{\hfil$#1$\hfil}} / \fi}

\begin{document}
\title{Nuclear Dependence in Weak Structure Functions and the Determination of Weak Mixing Angle}
\author{M. Sajjad Athar, H. Haider}
\address{Department of Physics, Aligarh Muslim University, Aligarh, India 202002}
\ead{sajathar@gmail.com}
\author{I Ruiz Simo, M.J. Vicente Vacas}
\address{Departamento de F\'\i sica Te\'orica e IFIC, Centro Mixto
Universidad de Valencia-CSIC, Institutos de Investigaci\'on de
Paterna, Apartado 22085, E-46071 Valencia, Spain}
%\ead{Manuel.J.Vicente@uv.es}
\begin{abstract}
We have studied nuclear medium effects in the weak structure functions $F^A_2(x)$ and $F^A_3(x)$ and in the extraction of weak mixing angle using Paschos Wolfenstein(PW) relation.
 We have modified the PW relation for nonisoscalar nuclear target. We have incorporated the 
medium effects like Pauli blocking, Fermi motion, nuclear binding energy, nucleon correlations, pion $\&$ rho cloud contributions, and shadowing and antishadowing effects. 
\end{abstract}
\section{Introduction}
In this paper, we have studied the impact of nuclear effects and nonisoscalarity corrections on the weak structure functions $F^A_2(x)$ and $F^A_3(x)$ 
using the expression discussed in the other paper presented in this conference. Using the results for $F^A_2(x)$ and $F^A_3(x)$, we obtain the 
results for charged and neutral current differential scattering cross sections~\cite{prc84,prc85}. The expression for the (anti)neutrino induced charged current 
differential scattering cross section is written as:
\begin{small}
\begin{eqnarray}\label{diff_dxdy_A}
\frac{d^2 \sigma^{\nu(\bar{\nu})A}}{d x d y}=
\frac{{G_F}^2ME_{\nu}}{\pi}
\left\{xy^2 F_1^{\nu(\bar{\nu})A} (x, Q^2)
+ \left(1-y-\frac{xyM}{2 E_{\nu}}\right) F_2^{\nu(\bar{\nu})A} (x, Q^2) \pm xy\left(1-\frac{y}{2}\right)F_3^{\nu(\bar{\nu})A} (x, Q^2)
\right\},
\end{eqnarray}
\end{small}
where the plus(minus) sign stands for the $\nu(\bar \nu)$ cross section. Similarly, one can write the differential scattering cross section for the
neutral current $\nu(\bar \nu)$ induced reactions by changing the couplings and the nucleon structure functions~\cite{prc}. Using these expressions, the result for 
the ratio 
\begin{small}
\begin{eqnarray}
 R^-&=&\frac{\frac{d\sigma_{NC}^{\nu A}}{dx\,dy}-\frac{d\sigma_{NC}^{\bar{\nu}A}}{dx\,dy}}{\frac{d\sigma_{CC}^
 {\nu A}}{dx\,dy}-\frac{d\sigma_{CC}^{\bar{\nu}A}}{dx\,dy}} \label{huma6}
\end{eqnarray}
\end{small}
is obtained. Paschos and Wolfenstein(PW)~\cite{Paschos} demonstrated that for an isoscalar nuclear target the ratio of neutral current to charged current cross sections is
related to the Weinberg angle $\theta_W$ as:
\begin{small}
\begin{eqnarray} \label{pwrelation}
R_{PW}=\frac{\sigma(\nu_\mu~N \rightarrow \nu_\mu~X)~-~\sigma(\bar\nu_\mu~N \rightarrow \bar\nu_\mu~X)}{\sigma(\nu_\mu~N \rightarrow \mu^-~X)~-~\sigma(\bar\nu_\mu~N \rightarrow \mu^+~X)}=\frac{1}{2}~-~\sin^2 \theta_W
\end{eqnarray}
\end{small}
\begin{figure}
\begin{center}
\includegraphics[height=4.8cm, width=7cm]{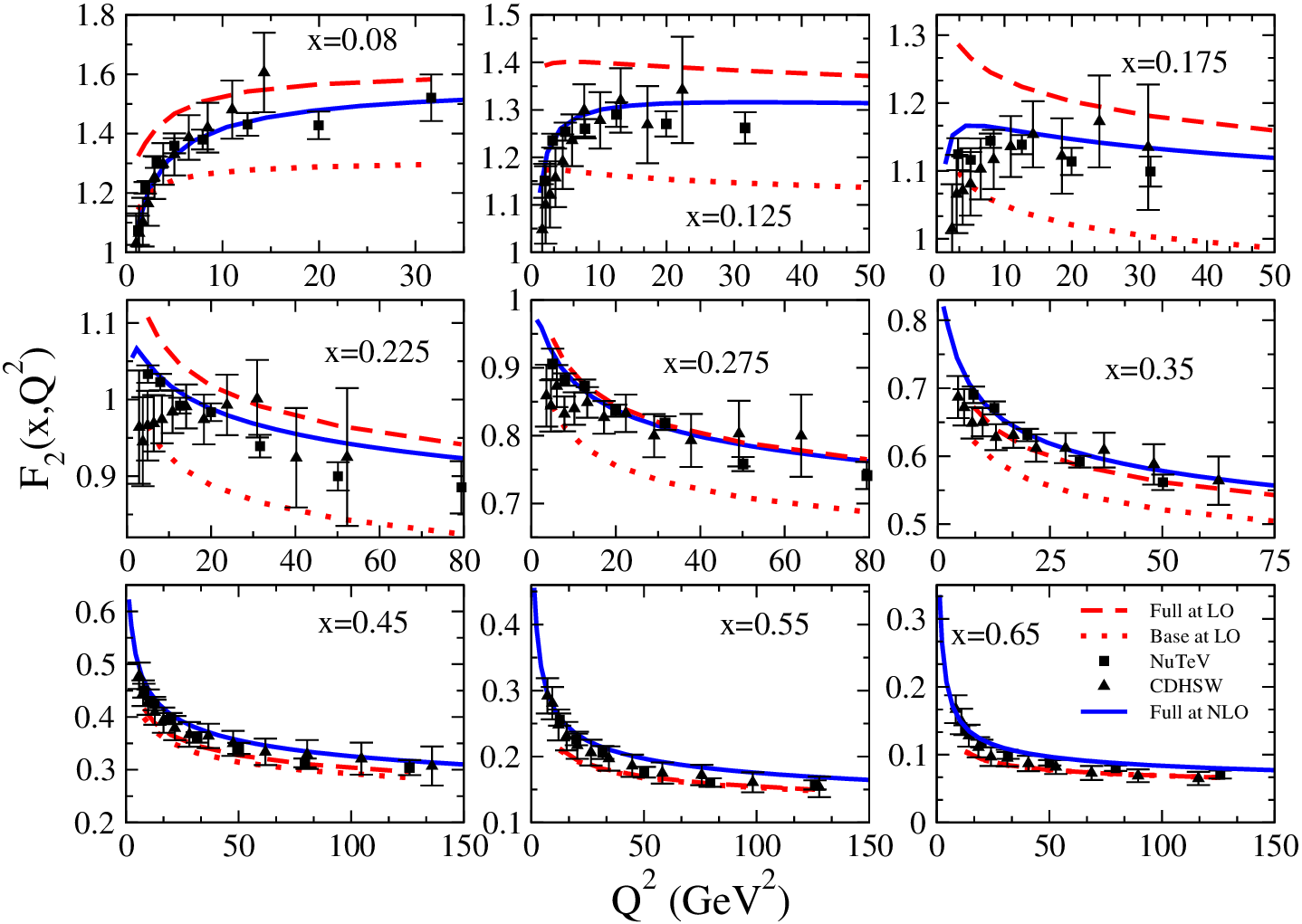}
\includegraphics[height=4.8cm, width=7cm]{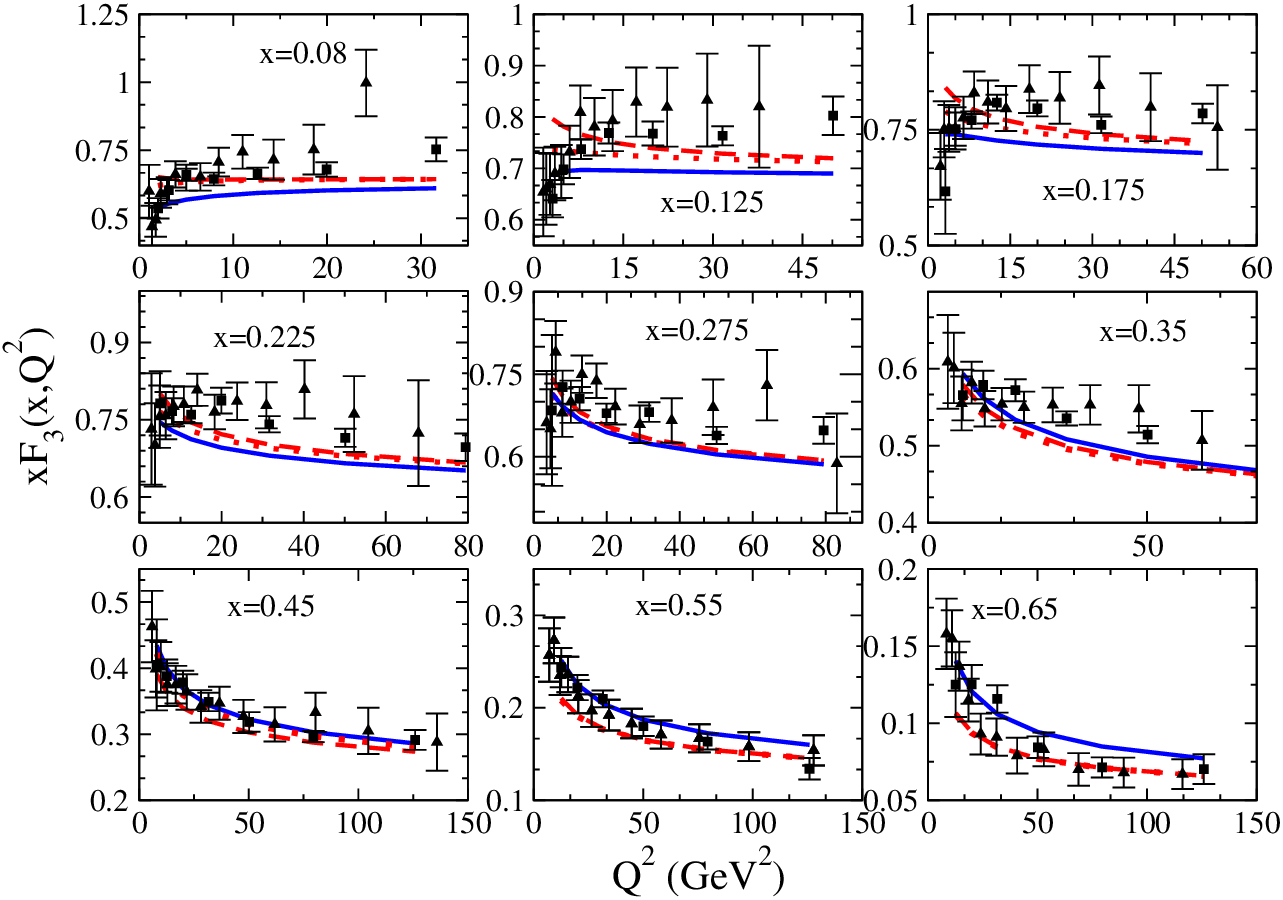}
\caption{Dotted line is our base result at LO for $F_i(x,Q^2)$ vs $Q^2$ in $^{56}$Fe(i=2(Left Panel), 3(Right panel)). 
Dashed(Solid) line is the full model at LO(NLO). The experimental points are from CDHSW~\cite{Berge} and NuTeV~\cite{Tzanov}.}
\label{fig_f2_f3_iron}
\end{center}
\end{figure}
The above relation is also valid for the ratio obtained using differential scattering cross sections under more general assumptions. NuTeV Collaboration~\cite{Zeller} has extracted the 
weak mixing angle using $\nu/\bar \nu$ beam on iron target and the above relation, and obtained $sin^{2}\theta_W=0.2277\pm0.0004$, which is
 3$\sigma$ above the global fit of $sin^2 \theta_W=0.2227\pm0.0004$ and is known as NuTeV anomaly.
Since iron is a nonisoscalar nuclear target, therefore, we have studied the effect of nuclear medium as well as nonisoscalarity correction on the extraction of 
weak mixing angle using PW relation. The details are given in Ref.~\cite{prc}. $R^-$ given by Eq.~\ref{huma6}, for a nonisoscalar nuclear target, may be written as
\begin{small}
\begin{eqnarray} \label{delta_modified}
 R^- &=&\frac{1}{2}-\sin^{2}\theta_W + \delta R^{-}
 \end{eqnarray}
 \end{small}
 where $\delta R^{-}$ is the nonisoscalarity effect the  expression for which is given in Ref.~\cite{prc}.
\begin{figure}
\begin{center}
\includegraphics[height=4.8cm, width=7cm]{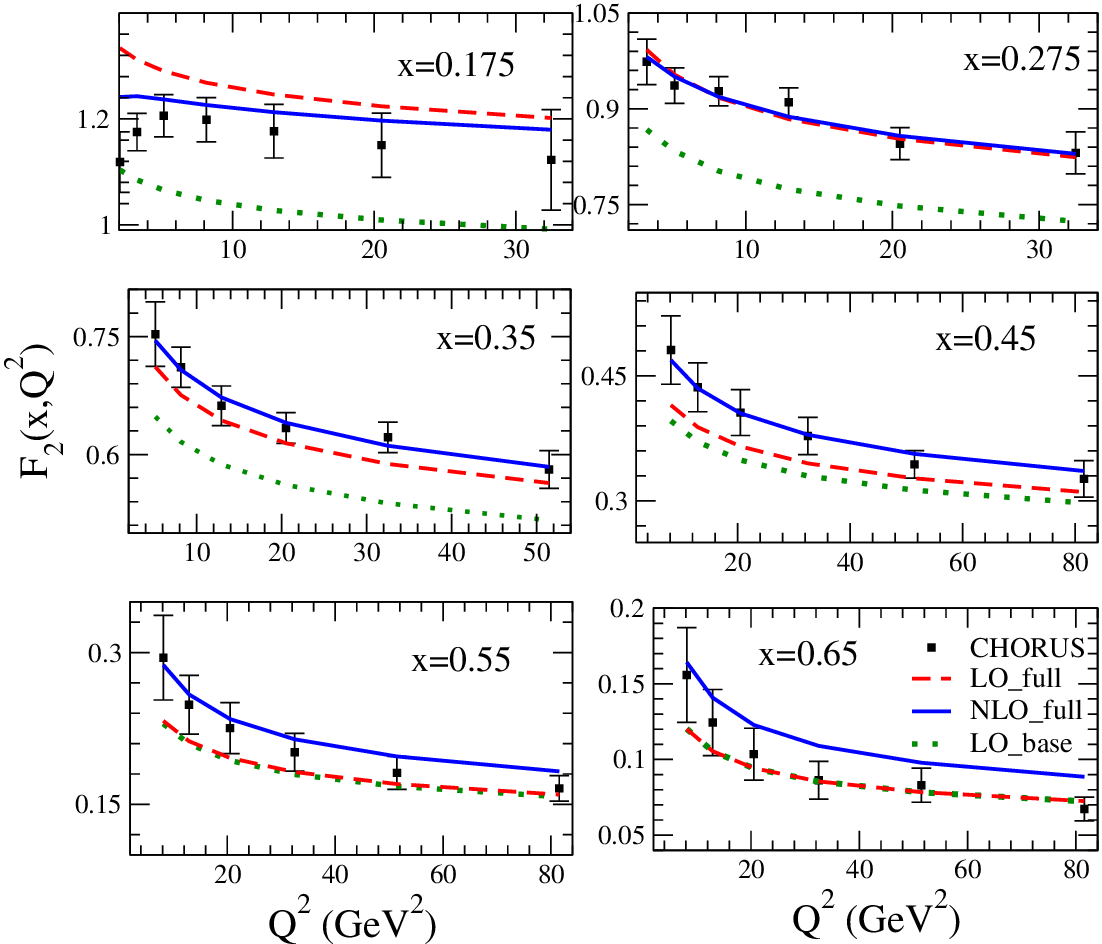}
\includegraphics[height=4.8cm, width=7cm]{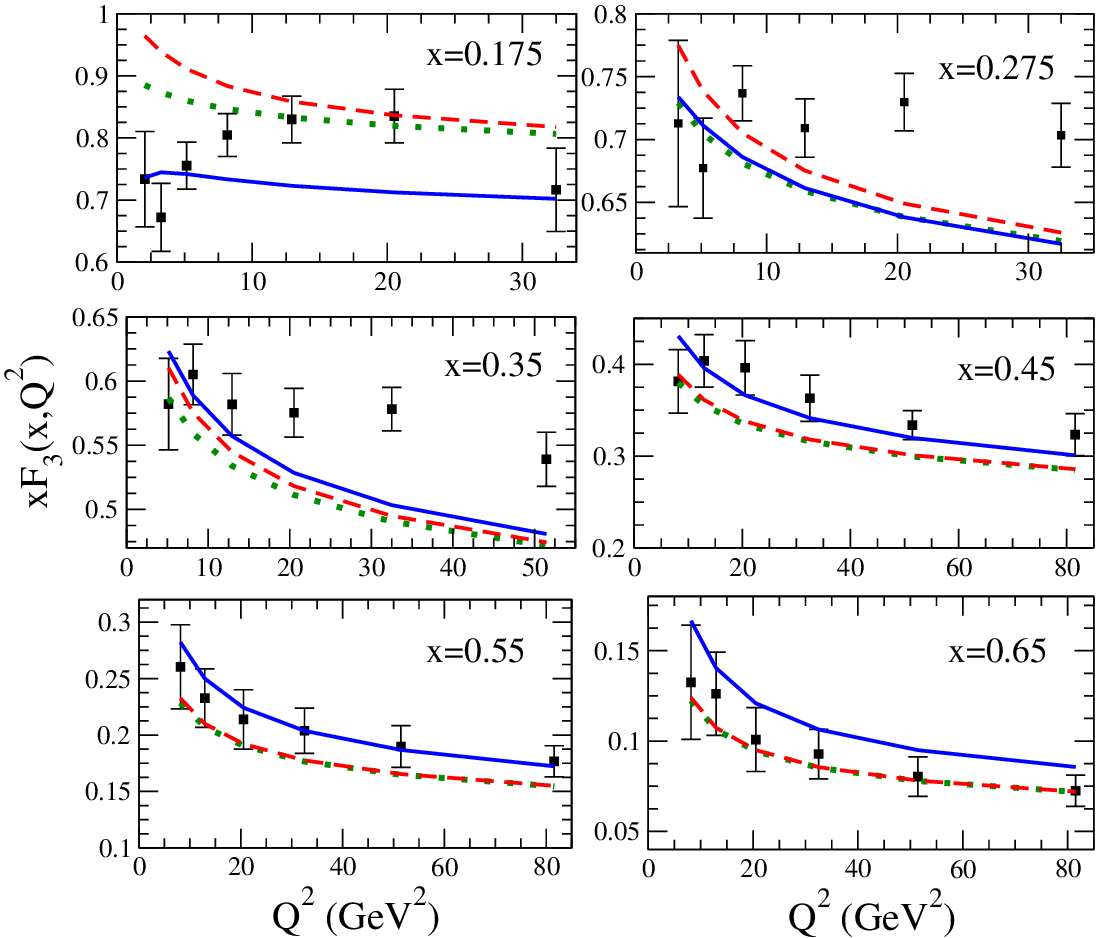}
\caption{Same results as in Fig.\ref{fig_f2_f3_iron} for $^{208}$Pb. The experimental points are from CHORUS~\cite{chorus1}.}
\label{fig_f2_f3_lead}
\end{center}
\end{figure}
\section{Results and Discussions}
\begin{figure}
\begin{center}
\includegraphics[height=5.0cm,width=12cm]{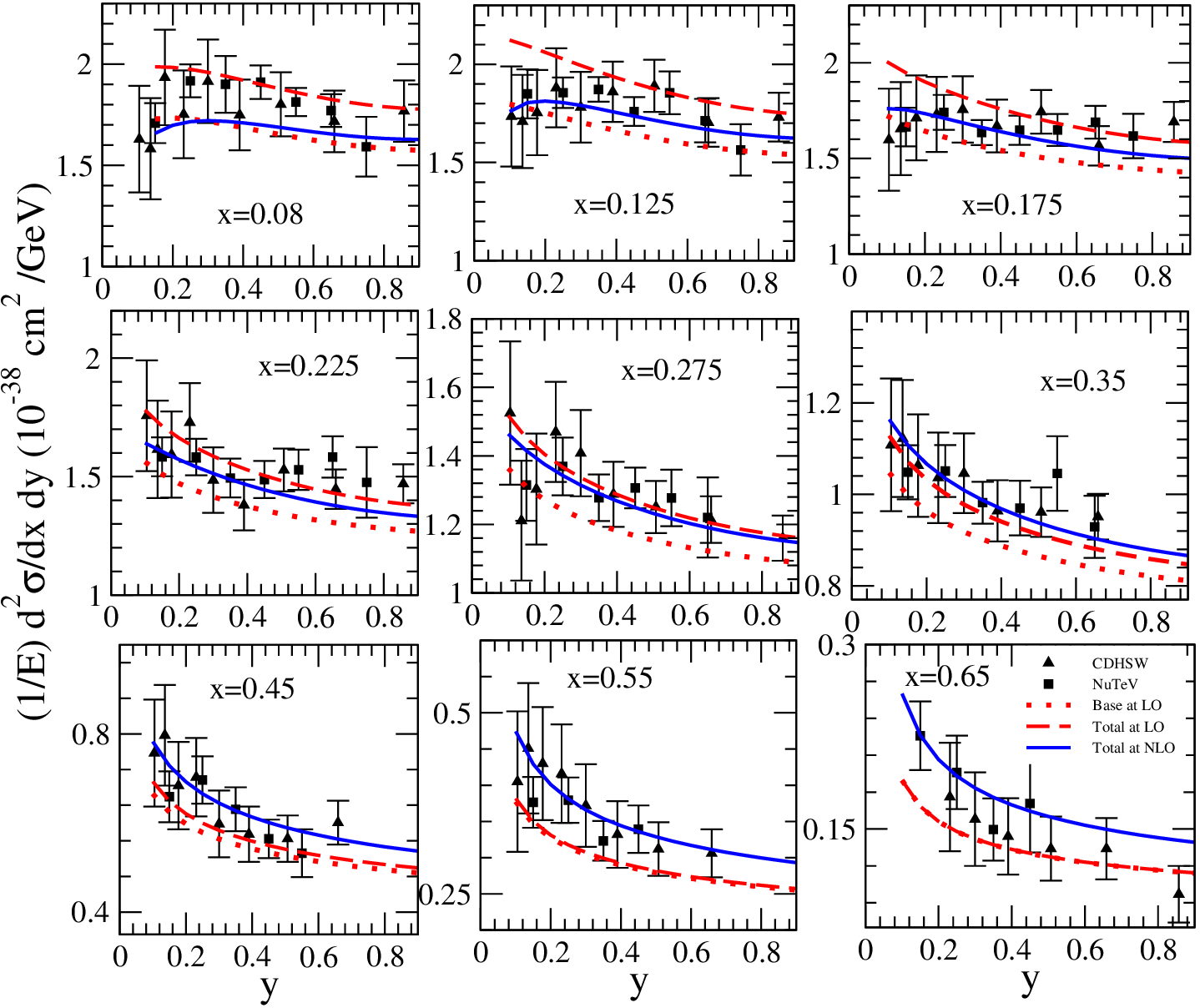}
\caption{$\frac{1}{E}\frac{d^2\sigma}{dxdy}$ vs y at different x for $\nu_\mu$($E_{\nu_\mu}=65$ GeV) induced reaction in $^{56}$Fe.
Dotted line is our base result at LO for $F_2(x,Q^2)$ vs $Q^2$ in $^{56}$Fe. Dashed(Solid) line is the full model at LO(NLO).
The experimental points are from CDHSW~\cite{Berge} and NuTeV~\cite{Tzanov}.}
\label{figFe}
\end{center}
\end{figure}
\begin{figure}
\begin{center}
\includegraphics[height=5.0cm,width=12cm]{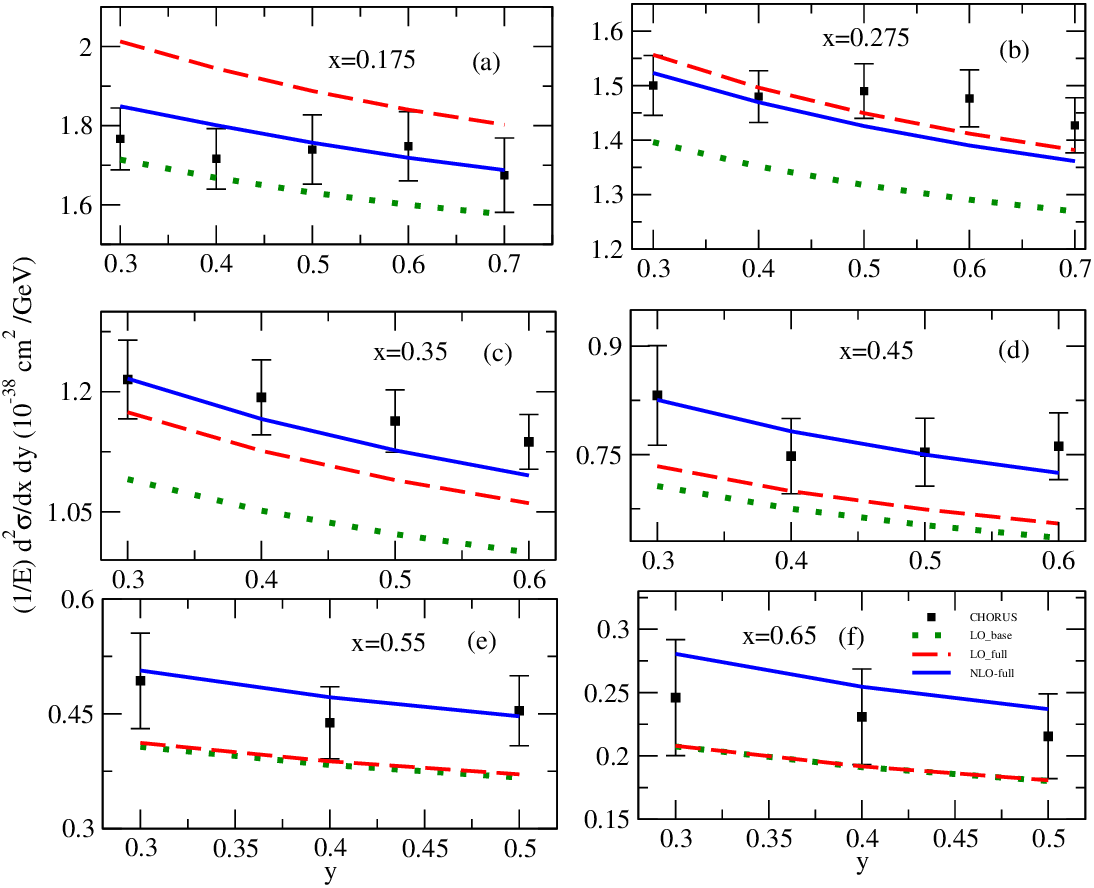}
\caption{Same results as in Fig.\ref{figFe} for $^{208}$Pb at $E_{\nu_\mu}=25$ GeV. The experimental points are from CHORUS\cite{chorus1}.}
\label{figPb}
\end{center}
\end{figure}
\begin{figure}
\begin{center}
\includegraphics[height=4.4cm,width=7cm]{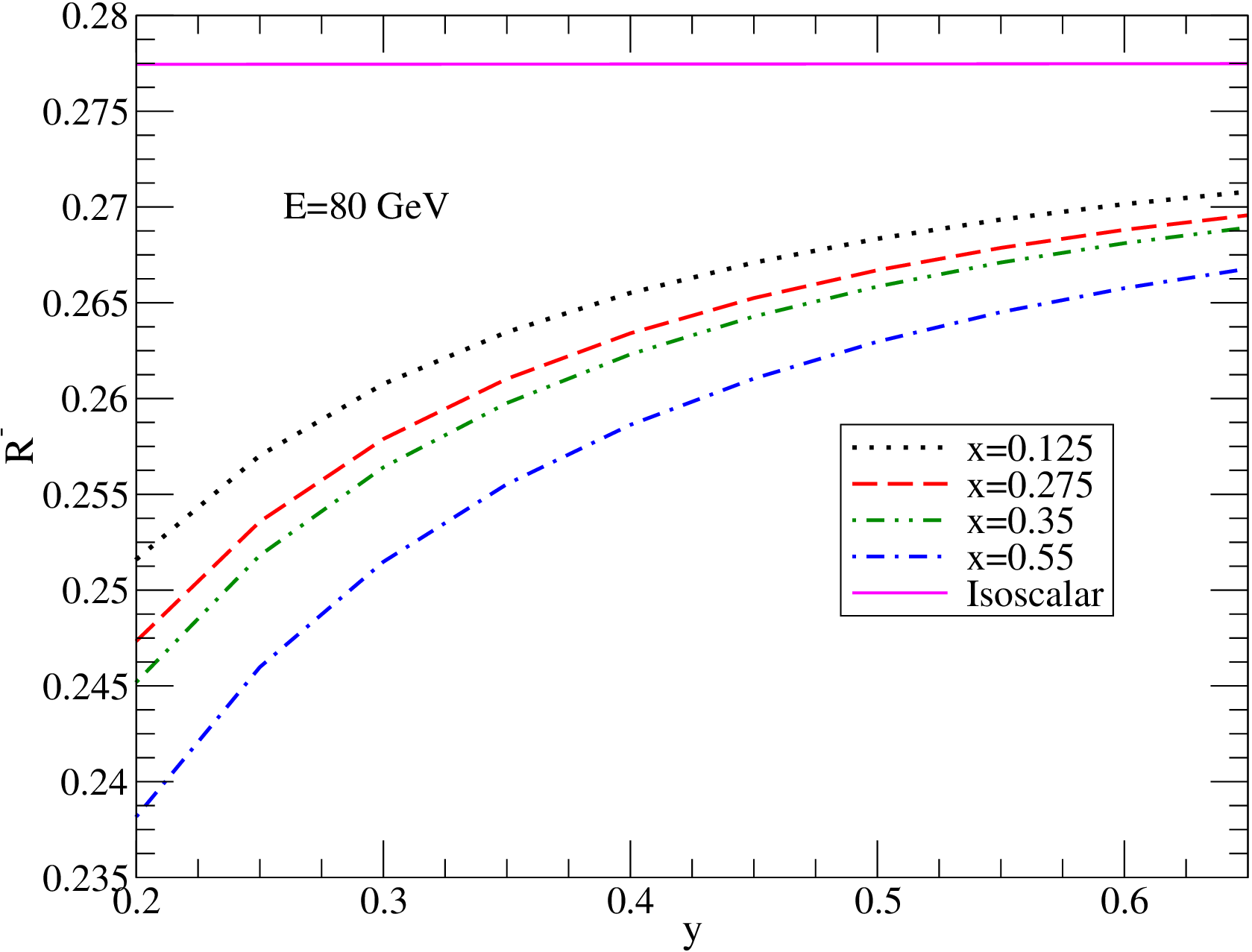}
\includegraphics[height=4.4cm,width=7cm]{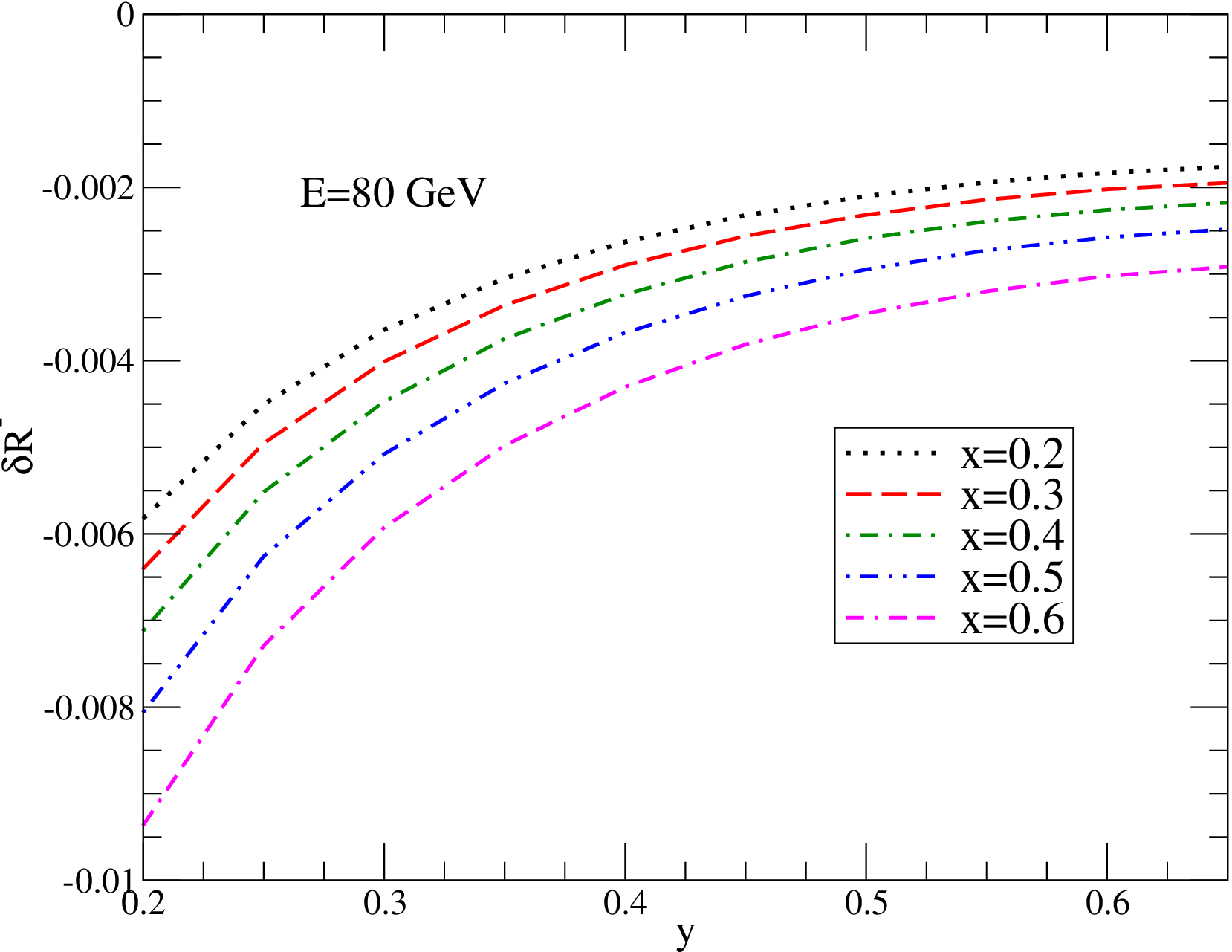}
\caption{$R^-$(Left panel) and $\delta R^{-}$(Right panel) as a function of y at different values of x.}
\label{rminusdelta}
\end{center}
\end{figure}
\begin{figure}
\begin{center}
\includegraphics[height=4.4cm,width=12cm]{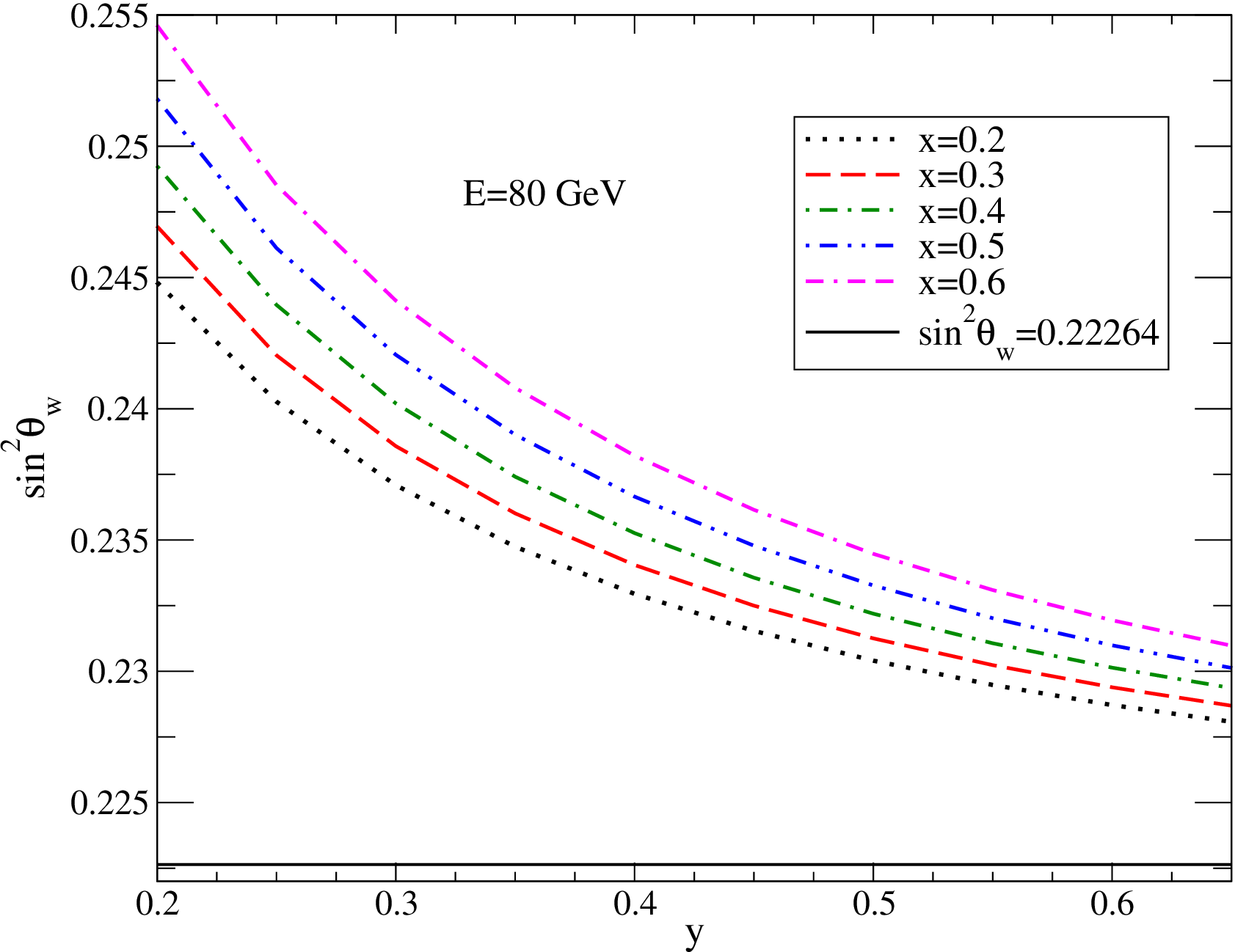}
\caption{$sin^{2}\theta_w$ vs y in $^{56}$Fe at different x treating it to be nonisoscalar nuclear target.}
\label{sinthetafig}
\end{center}
\end{figure}

Here we present the numerical results for $F^A_2$ and $F^A_3$ structure functions in Fig.\ref{fig_f2_f3_iron} for iron nucleus along with the experimental data of 
 CDHSW~\cite{Berge} and NuTeV~\cite{Tzanov} collaborations for a wide range of x and $Q^2$. The results are obtained using target mass correction and for nucleon parton distribution
  functions we have used CTEQ PDFs. When the effect of 
 Pauli blocking, Fermi motion, nuclear binding energy and nucleon correlations are taken, we call the results as the base results. Then we include pion, rho cloud contributions
  and shadowing and antishadowing effects and the results obtained using the full calculation, we call it as the results with the full calculation. We find that the effect of shadowing is
 about 3-5$\%$ at x=0.1 for Q$^2$=1-5GeV$^2$ and 1-2$\%$ at x=0.2 for Q$^2$=1-5GeV$^2$ which dies out with the increase in x and Q$^2$.
Furthermore, in the evaluation of $F^A_2$ nuclear structure function there is contribution from pion and rho clouds. 
Pion contribution 
is significant in the region of $0.1~<~x~<~0.4$ which is $\approx$ 14-16$\%$ at x=0.1 which reduces to 4-6$\%$ at mid values of x.
It is the meson cloud contribution which is dominant at low and intermediate x for $F_2$. 
In Fig.\ref{fig_f2_f3_lead}, we have shown the numerical results for $F^A_2$ and $F^A_3$ structure functions in lead along with the experimental data of 
CHORUS collaboration for a wide range of x and $Q^2$. Similar to the case of iron, shadowing is negligible as compared to the other nuclear effects 
and meson cloud contribution plays a significant role at low and mid x.
We observe that the results at NLO are in better agreement with experimental data.
In Fig.\ref{figFe}, we have shown the results for $\frac{1}{E}\frac{d^2\sigma}{dxdy}$ in $^{56}$Fe at $E_{\nu_\mu}$=65 GeV. 
Similarly in Fig.\ref{figPb}, we have shown the results for $\frac{1}{E}\frac{d^2\sigma}{dxdy}$ $^{208}$Pb at $E_{\nu_\mu}$=25 GeV. 
We find that the results of the full calculation at NLO are in general in good agreement with the experimental observations of CDHSW, NuTeV and CHORUS collaborations.
In Fig. \ref{rminusdelta}, we have presented the results for $R^-$ vs y and $\delta R^-$ vs y for different values of x at  $\nu(\bar \nu)$ energy E= 80 GeV in the left(right)
 panel respectively.
We find that $R^-$ is almost independent of x and y for an isoscalar target, while for the  nonisoscalar target there is x as well as y dependence.
In the right panel of Fig.\ref{rminusdelta}, we find that the effect of non-isoscalarity is large at low y and high x which decreases with the increase in the value of
y. In Fig.\ref{sinthetafig}, we have depicted the results of $sin^{2}\theta_W$ vs y for different values of x at $\nu(\bar \nu)$ energy E= 80 GeV.
 We find that due to medium effects, $sin^{2}\theta_W$ is different from the global fit,
 and this difference is $\approx 7\%$ when evaluated for low value of y at  x=0.2 and this decreases to $1\%$ at high values of y, while this change is $\approx 9\%$ when
  calculated for low y at x=0.6 and this reduces to $2\%$ at high values of y.
  Thus, we observe that nonisoscalarity as well as nuclear medium effects are important while extracting  $sin^{2}\theta_W$.

One of the authors(MSA) is thankful to PURSE program of D.S.T., Govt. of India and the Aligarh Muslim University for the financial support. This research was supported by the Spanish Ministerio de Economía y Competitividad and
European FEDER funds under Contracts FIS2011-28853-C02-01, by Generalitat Valenciana under Contract No.
PROMETEO/20090090 and by the EU HadronPhysics3 project, Grant Agreement No. 283286.

\end{document}